\documentclass[12pt]{article}
\usepackage{graphicx}
\usepackage[square]{natbib}
\usepackage{amsmath,amssymb}
\usepackage{rotating}
\usepackage[footnotesize]{caption}
\title{\Huge On the dynamics \\ of \\ Swimming Linked Bodies
\vskip 3cm }
\author{ \Large  {J. B. Kajtar \quad and \quad J. J. Monaghan }  \\
 { \small School of Mathematical Sciences } \\ {\small Monash University,  Vic 3800 Australia.}\\
{ \small email: joe.monaghan@sci.monash.edu.au }}\date {\small 24 July 2009}

\begin{document}

\maketitle
\vskip3cm

\newpage
\thispagestyle{empty}

\begin{abstract}  
In this paper we study the motion of three linked ellipses moving through a viscous fluid in two dimensions.  The angles between the ellipses  change with time in a specified manner (the gait) and the resulting time varying configuration is similar to the appearance of a swimming leech.   We simulate the motion using the particle method Smoothed Particle Hydrodynamics (SPH) which we test by convergence studies and by comparison with the inviscid results of Kanso et al. (2005) and the viscous results of Eldredge (2006, 2007, 2008). We determine how the average speed and power output depends on the amplitude and oscillation frequency of the gait.  We find that the results fit simple scaling rules which can related to  the analytical results of G.I. Taylor for the swimming of long narrow animals (1952).  We apply our results  to estimate the speed of a  swimming leech with reasonable accuracy, and we determine the minimum power required to propel the bodies at a specified average speed. 

\end{abstract}
\newpage


\section{Introduction}

The subject of this paper is the motion of linked rigid bodies moving in a weakly compressible, viscous  fluid.  It is closely connected with mathematical and computational studies of the swimming of fish, and with the motion of underwater vehicles and robotic fish propelled by changes of shape.  Our approach is primarily computational using the SPH algorithms of Kajtar and Monaghan (2008) to establish scaling relations for the motion. The work is closely related to recent work on the motion of linked bodies in an infinite, two dimensional fluid which may be inviscid (Kanso et al., 2005, Melli et al., 2006) or viscous (Eldredge, 2006, 2007, 2008).  When the fluid is inviscid it is  possible to bring powerful mathematical formalisms to bear on the problem in a manner similar to the motion of a single body in an inviscid fluid (see for example Lamb, 1932).  However, for problems involving free surfaces, or complicated rigid boundaries, or a stratified fluid, these methods become very complicated. Our approach  is capable of handling arbitrary body shapes and boundaries though in the present paper we concentrate on motion of linked ellipses moving in a periodic domain.  The SPH method also  has advantages over the vortex particle method of Eldredge (2006) for problems where the bodies penetrate a free surface, but in the present case no such difficulties exist and the vortex particle method provides a convenient comparison for the SPH calculations.
  
The bodies we consider are solid bodies linked by virtual rods  which join at pivot points. The rods are described as virtual because they do not have any mechanical function except to define the direction of  fixed lines in the bodies. In particular, fluid can flow between the ellipses. The angles between the rods (and therefore the bodies) are specified as an oscillating function of time.  A specification of the time variation of these angles is called the gait.
  
Our aim is to determine the scaling relations which relate the speed and power developed by the linked bodies to the frequency and amplitude of a standard gait which propels the linked bodies along a  path which is, on average,  a straight line.  A related problem was considered by Taylor (1952) who studied the motion of a long slender body and applied his results to the motion of a leech and a snake moving in water.  Our three ellipse system has a motion which is similar to that of the leech and snake because their oscillations are are roughly sinusoidal,  and are therefore not too different from the oscillations of our connected ellipses. Taylor's analysis provides a remarkably accurate guide for the functional form of the dependence of the velocity and power of our three ellipse system  on the frequency and amplitude of the gait.

The plan of the paper is to first discuss the SPH algorithm. We then show by convergence studies that a periodic domain can be used to represent the infinite domain with errors that are typically $5 \%$.  We also establish convergence with resolution.  We first compare our results with the viscous results of Eldredge (2008) for massless bodies by a series of simulations with decreasing body mass.  The agreement is very satisfactory.  We then compare our results to the inviscid results of Kanso et al. 2005 by changing the viscosity so that the Reynolds number varies from 50 to 5000.  This comparison shows that the SPH results converge to the inviscid results for the highest Reynolds number. We  then discuss scaling relations for the velocity and power output, and relate them to Taylor's (1952) analytical relations for the velocity and power output of long narrow animals swimming.  We apply these results to the swimming of a leech. Finally we determine the minimum power, and corresponding gait, required to propel the bodies at a specified average speed.  Throughout this paper we use SI units.
   
\section{Equations of motion and constraints}
\subsection{Equations of motion}
\begin{figure}[htbp]
\begin{center}
\includegraphics[width= 0.9 \textwidth]{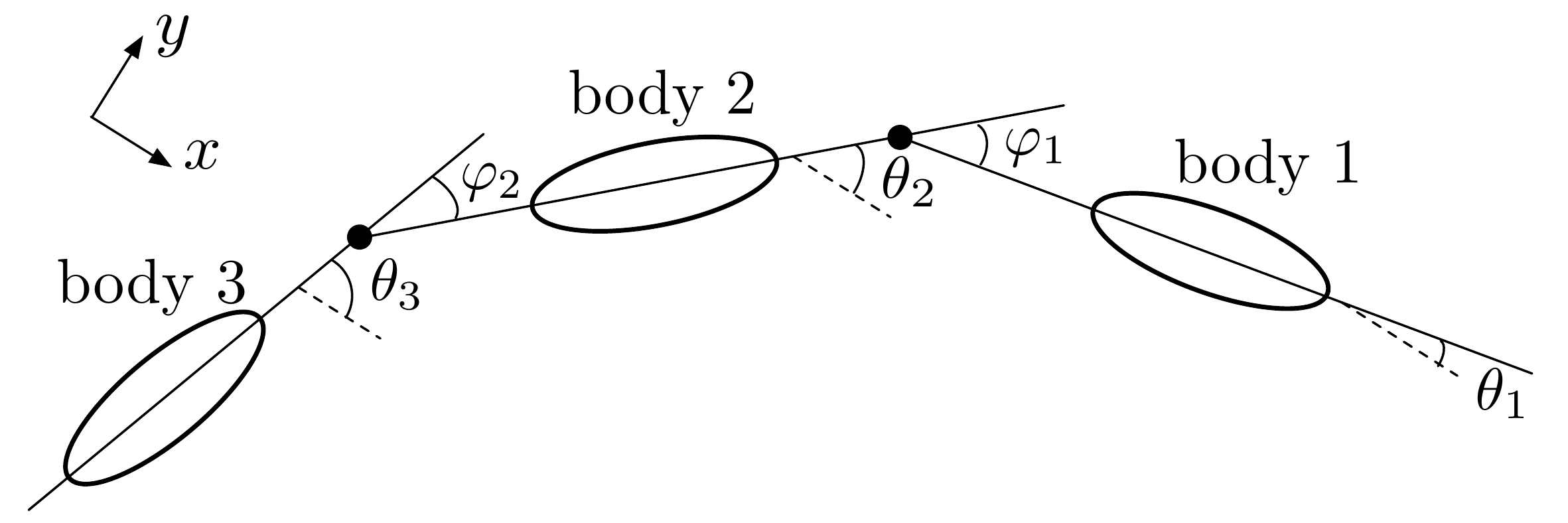}
\caption{The configuration of the bodies (assumed to be ellipses). The link position is denoted by a filled circle. The straight line through a body passes through its centre of  mass and is assumed to be rigidly attached to the body with one end attached to the link.  The angles $\theta$ are defined relative to a fixed direction in space (shown by parallel dotted lines) which is taken to be the $x$ axis of a cartesian coordinate system in our calculations.  The angles $\varphi$ determine the gait and are specified functions of  time.}
\label{fig:bodyangles}
\end{center}
\end{figure}

\setcounter{equation}{0}
We consider motion in two dimensions and use cartesian coordinates. A typical configuration of the bodies is shown in Figure \ref{fig:bodyangles}. The motion of the fluid, which is assumed incompressible, is specified by the Navier Stokes equations. In cartesian tensor form these equations are
\begin{equation}\label{eqn:navierstokes}
\frac{d v_i}{dt} =  \frac{1}{\rho} \frac{\partial \sigma_{ij} }{\partial x^j} + \frac{1}{\rho} \sum_{k =1}^{N_b} \sigma_{ij} n_j(k) \delta(s_k),
\end{equation}
where $N_b$ denotes the number of bodies, $\sigma_{ij}$ is the stress tensor 
\begin{equation}\label{eqn:stresstensor}
\sigma_{ij} = - P \delta_{ij} + \mu \left ( \frac{\partial v_i}{\partial x_j} +  \frac{\partial v_j}{\partial x_i} - \frac23 \delta_{ij}  \frac{\partial v_\ell}{\partial x_\ell} \right ),
\end{equation}
$P$ is the pressure and $\mu$ the shear viscosity coefficient. The function $\delta(s_k)$ is a one dimensional delta function, and $s_k $ is the perpendicular distance from  the surface $A_k$ of body $k$ to the position where the fluid acceleration is required.  The unit vector $n_j(k)$ is directed from body $k$ into the fluid. The introduction of forces into the acceleration equation (\ref{eqn:navierstokes}) as an alternative to specifying boundary conditions on the velocity is due to Sirovich (1967, 1968). In his formulation, as in ours, $\delta(s)$ is a delta function defined  so that for any quantity $B({\bf r})$ 
\begin{equation}
\int B \delta(s) d{\bf r} =  \int B dA,
\end{equation}
where the first integral is over the volume and the second integral is over the surface. In this way a volume integral involving a delta function becomes equivalent to a surface integral, and the body force per unit volume in (\ref{eqn:navierstokes}) becomes a force per unit area. This force provides both the pressure which prevents penetration of the rigid body, and the viscous traction term.  It mimics the fundamental molecular basis of the boundary conditions namely that the atoms of the fluid do not penetrate the atoms of the solid because of the atomic forces between the liquid and the solid atoms.

A closely related method of using boundary forces is due to Peskin (1977) who simulated elastic membranes such as the heart interacting with a fluid.  Peskin's equations (2.3) to (2.6) are essentially those of Sirovich, though the Peskin deals with an elastic material and Sirovich  assumes the body is rigid.  Further details about Peskin's formulation can be found in Peskin (2002). 
 
We denote an element of area on the surface of body $k$  by $dA(k)$.  The motion of the centre of mass ${\bf R}(k)$ of solid body $k$ (with mass $M(k)$) is given by 
\begin{equation}
M(k) \frac{ d^2  R^i(k)}{dt^2} = - \int \sigma^{ij} n^j(k) dA(k) + F^i(k),
\end{equation}
where ${\bf F}(k)$ is the force due to the constraints. The rotation of rigid body $k$, with moment of inertia $I(k)$), is given by
\begin{equation}
I(k) \frac{d ^2\theta_k}{dt^2} =  \int ( {\bf d}_k \times {\bf b} )dA(k) + \tau(k),
\end{equation}
where ${\bf d}_k$ is a vector from the centre of mass of body $k$ to the element of area $dA_k$, ${\bf b}$ is the force on the element of area, and $\tau(k$) is the constraint torque on body $k$.

In the following, to simplify the notation, the subscript $k$ will always denote the label of a body. Thus, for example, ${\bf R}(k)$ will be replaced by ${\bf R}_k$. 
\subsection{Constraint equations forces and torques}

The angle $\theta_k$ which fixes the rotation of body $k$ is defined as the positive rotation of a line fixed in the body from the $x$ axis of a cartesian coordinate system fixed in space.  For simplicity we assume the line fixed in the body is an axis of symmetry. The constraint conditions on the angles are 
\begin{equation}
\varphi_m = \theta_{m+1} - \theta_m,
\end{equation}
where $m$ is the link number and $\varphi_m$ is a specified function.  The form of the $\varphi_m$ determines the gait of the bodies.  For the examples we consider here there are three bodies and two links as shown in Figure \ref{fig:bodyangles}.  In the simplest case $\varphi_m$ is a function of $t$ but, in general, it depends on other variables.  For example, in a biological problem, it could depend on the centre of mass coordinates in such a way that the fish slows down when it enters a region where food is abundant.  

In addition to the constraints on the angles there are constraints associated with the links.  We assume the link, or pivot, is  at a distance $\ell_k$ from the centre of mass of body $k$.  The condition on the $X$ components of the centres of mass of bodies $k$ and $k+1$ is that the $X$ coordinate of the link between them is given by
\begin{equation}
X_k  - \ell_k  \cos{(\theta_k)}  = X_{k+1} +  \ell_{k+1} \cos{(\theta_{k+1})},
\end{equation}
or 
\begin{equation}\label{eqn:xcons}
X_k  - \ell_k  \cos{(\theta_k)}  -X_{k+1} -  \ell_{k+1} \cos{(\theta_{k+1})} = 0.
\end{equation}
Similarly the $Y$ constraint is 
\begin{equation}\label{eqn:ycons}
Y_k  - \ell_k  \sin{(\theta_k)}  -Y_{k+1} - \ell_{k+1} \sin{(\theta_{k+1})} = 0.
\end{equation}
These constraints enable the coordinates of the centres of mass of the bodies, and their angles $\theta$ to be written in terms of those of any selected body.  Similarly, by differentiating the constraint conditions with respect to time, the velocities $\dot X$ and $\dot Y$ and angular velocity $\Omega$ of the bodies can  be written as functions of the same selected body.  The number of degrees of freedom (coordinates and velocities) of $N$ linked bodies in two dimensions is therefore 6 compared with the $6N$ degrees of freedom of $N$ independent bodies in two dimensions.  If the $\varphi_m$ are functions of $t$ alone it is possible to reduce the equations of motion to those involving the coordinates and velocities of one of the bodies.  This can also be done when the $\varphi_m$ are functions of both coordinates and time but it is inconvenient to eliminate variables and, in our view, simpler to take account of the constraints by using Lagrange multipliers. For that reason we use Lagrange multipliers even though, in the applications to be described in this paper, the $\varphi_m$ are functions of $t$ only. 
 
For the case of three bodies we have two links and therefore 6 constraints.  We denote the Lagrange multipliers for the $X$, $Y$ and $\theta$ constraints of link $m$ by  $\lambda_X^{(m)}$, $\lambda_Y^{(m)}$ and $\lambda_\theta^{(m)}$ respectively.  Using standard methods  for holonomic constraints (e.g. Landau and Lifshitz, 1976) we find the following expressions for the constraint forces ${\bf F}_k$ and torques $\tau_k$ for the various bodies. For body 1
\begin{equation}\label{eqn:cforce1}
{\bf F}_1 = (\lambda_X^{(1)} , \lambda_Y^{(1)}),
\end{equation}
for body 2
\begin{equation}\label{eqn:cforce2}
{\bf F}_2 = (-\lambda_X^{(1)} , -\lambda_Y^{(1)}) + (\lambda_X^{(2)},\lambda_Y^{(2)}),
\end{equation}
and for body 3
\begin{equation}\label{eqn:cforce3}
{\bf F}_3 =  (-\lambda_X^{(2)},-\lambda_Y^{(2)}).
\end{equation}
These constraint forces do not affect the total linear momentum of the bodies because they sum to zero. 

The constraint torque on body 1 is 
\begin{equation}\label{eqn:ctorque1}
\tau_1 = -\lambda_{\theta}^{(1)}+ \lambda_X^{(1)} \ell_1 \sin{(\theta_1)} - \lambda_Y^{(1)} \ell_1 \cos{(\theta_1)},
\end{equation}
on body 2 it is
\begin{equation}\label{eqn:ctorque2}
\tau_2 = \lambda_{\theta}^{(1)} - \lambda_{\theta}^{(2)}+ \left( \lambda_X^{(1)}  + \lambda_X^{(2)} \right) \ell_2  \sin{(\theta_2)}- \left( \lambda_Y^{(1)}+ \lambda_Y^{(2)} \right) \ell_2 \cos{(\theta_2)}
\end{equation}
and on body 3 it is
\begin{equation}\label{eqn:ctorque3}
\tau_3 =  \lambda_{\theta}^2+ \lambda_X^{(2)} \ell_3 \sin{(\theta_3)} - \lambda_Y^{(2)} \ell_3 \cos{(\theta_3)} .
\end{equation}

The constraint forces and torques are provided by the engines which drive the angular variation between the bodies. In the case of fish these engines are the muscles of their bodies and the work done is provided by the internal chemical energy generated by the fish. The way these constraint forces and torques affect the angular momentum will be discussed in $\S 5$. 

\section{SPH equations for the fluid}
\setcounter{equation}{0}
The form of the SPH equations that we use is discussed in more detail by Monaghan (1992, 2005). For the liquid SPH particles the acceleration equation is 
\begin{equation}\label{eqn:sphaccel}
\frac{ d {\bf v}_a}{dt} = - \sum_b m_b \left ( \frac{P_a}{\rho_a^2 }  +\frac{P_b}{\rho_b^2} + \Pi_{ab} \right ) \nabla_a W_{ab} + \sum_{k=1}^{N_b} \sum_{j\in S_k} \left [ {\bf f}_{aj} - m_j\Pi_{aj} \nabla_a W_{aj} \right ].
\end{equation}   
In this equation the mass, position, velocity, density, and pressure of particle $a$ are $m_a$, ${\bf r}_a$, ${\bf v}_a$,  $\rho_a$, and $P_a$  respectively.  $W_{ab}$ denotes the smoothing kernel $W({\bf r}_a - {\bf r}_b, \bar h_{ab})$ and $\nabla_a$ denotes the gradient taken with respect to the coordinates of particle $a$. In this paper $W$ is the fourth degree Wendland kernel for two dimensions (Wendland, 1995), and has support $2\bar h_{ab}$.  In the present calculations the $\bar h_{ab}$ used in $W_{ab}$ is an average $\bar h_{ab}= (h_a + h_b)/2$.  The choice of $h$ is discussed in detail by Monaghan (1992, 2005). In this paper we choose $h$ to be 1.5 times the initial particle spacing so that the interaction between any two fluid particles is zero beyond 3 initial particle spacings. 

The first summation in (\ref{eqn:sphaccel}) is over all fluid particles and is the SPH equivalent of the first term on the right hand side of (\ref{eqn:navierstokes}). The last term in (\ref{eqn:sphaccel}) is the contribution to the force per unit mass on fluid particle $a$ due to boundary particles and is equivalent to the last term in (\ref{eqn:navierstokes}).  A body label is denoted by $k$, and $j \in S_k$ is one of the set of  boundary particle labels on body $k$. The term ${\bf f}_{aj}$ is the non-viscous boundary particle force per unit mass on fluid particle $a$ due to boundary particle $j$.  In the present paper we use the boundary forces analysed by Monaghan and Kajtar (2009).  The force ${\bf f}_{aj}$ acts on the line joining particle $a$ and $j$.  The boundary particles delineate the boundaries, and produce forces on the fluid in a similar manner to the delta function forces of Sirovich discussed after (\ref{eqn:stresstensor}).

 The viscosity is determined by $\Pi_{ab}$ for which we choose the form (Monaghan 1997, 2005)
\begin{equation}
\Pi_{ab} = -\frac{ \alpha v_{sig} {\bf v}_{ab} \cdot {\bf r}_{ab}   }{\rho_{ab} |{\bf r}_{ab}|}.
\end{equation}
In this expression $\alpha $ is a constant, and the notation ${\bf v}_{ab} = {\bf v}_a - {\bf v}_b$ is used. $\rho_{ab}$ denotes the average density $\frac12(\rho_a + \rho_b)$.  We take the  signal velocity to be 
\begin{equation}
v_{sig} = \frac12( c_a + c_b) -  2\frac{ {\bf v}_{ab} \cdot {\bf r}_{ab}}{r_{ab} },
\end{equation} 
where $c_a$ is the speed of sound at particle $a$ (Monaghan, 1997, though here we take $v_{sig}$ to be half used in that paper and $\alpha$ is therefore a factor 2 larger).  
$v_{sig}$ is dominated by the terms involving the speed of sound. The kinematic viscosity can be estimated by taking the continuum limit which is equivalent to letting the number of particles go to infinity while keeping the resolution length $h$ constant.  By a calculation similar to that  in Monaghan (2005) it is found that the kinematic viscosity for the Wendland kernel is given by
\begin{equation}
\nu = \frac{1}{8} \alpha h v_{sig}.
\end{equation}
SPH calculations for shear flow agree very closely with theoretical results using this kinematic viscosity (Monaghan, 2006).  

The pressure is given by
\begin{equation}
P_a = \frac{ \rho_0 c_a^2}{7} \left (  \left(  \frac{\rho_a}{\rho_0} \right )^7 -1   \right ),
\end{equation}
where $\rho_0$ is the reference density of the fluid. To ensure the flow has a sufficiently low Mach number to approximate  a constant density fluid accurately, we determine the speed of sound by $c_a \sim 10 V$ where $V$ is the maximum speed of the fluid relative to the bodies. In this case $v_{sig}$ is dominated by the first two terms. The precise value of $c_a$ will be specified for each simulation. 

The form of the SPH continuity equation we use here is 
\begin{equation}
\frac{d\rho_a}{dt} = \sum_b m_b {\bf v}_{ab} \cdot \nabla W_{ab},
\end{equation}
and the position of any fluid  particle $a$ is found by integrating
\begin{equation}
\frac{ d{\bf r}_a}{dt} = {\bf v}_a.
\end{equation}

In the present simulations the liquid SPH particles were initially placed on a grid of squares and thereafter allowed to move in response to the forces.  The time stepping of the SPH equations uses an algorithm which is symplectic in the absence of dissipation.  The details of this scheme are given by Kajtar and Monaghan (2008).
\section{SPH equations for the rigid bodies}
\setcounter{equation}{0}
 
The non-viscous force on boundary particle $j$ due to all fluid particles  is 
\begin{equation}
{\bf f}_j^{(nv)} = m_j\sum_a  {\bf f}_{ja},  
\end{equation}  
where ${\bf f}_{ja}$ is the force per unit mass on boundary particle $j$ due to fluid particle $a$. The viscous force is 
\begin{equation}
{\bf f}^{(v)}_j = -m_j \sum_a m_a \Pi_{aj} \nabla_j W_{aj}  = m_j \sum_a m_a \Pi_{aj} \nabla_a W_{aj},
\end{equation}
where we have used the fact that $\nabla_j W_{aj} = -\nabla_a W_{aj}$. The total force on particle $j$ is
\begin{equation}
{\bf f}_j =  {\bf f}_j^{(nv)} + {\bf f}^{(v)}_j .
\end{equation}

The equation for the centre of mass motion of body $k$ is then
\begin{equation}
M_k \frac{ d {\bf V}_k}{dt} = \sum_{j \in S_k} {\bf f}_j +   {\bf F}_k,
\end{equation}
and the torque equation is 
\begin{equation}
I_k \frac{ d \Omega_k}{dt} = \sum_{j \in S_k} ({\bf r}_j - {\bf R}_k) \times{\bf f}_j + \tau_k.
\end{equation}
  
The motion of a boundary particle can be determined from the motion of centre of mass and the rotation about the centre of mass. Thus for particle $j$ on body $k$,
\begin{equation}
\frac{  d{\bf r}_j}{dt} = {\bf V}_k  + \Omega_k \hat{\bf z} \times ( {\bf r}_j - {\bf R}_k),
\end{equation}
where, in this two dimensional problem, the rotation is around the $z$ axis which is perpendicular to the plane of the motion. 

\section{Conservation of linear and angular momentum}
\setcounter{equation}{0}
The total rate of change of the linear momentum of the rigid bodies with respect to time is 
\begin{equation}
\sum_k M_k \frac{d {\bf V}_k}{dt} = \sum_k \sum_{j \in S_k} {\bf f}_j = \sum_k \sum_{j \in S_k} \sum_a m_j\left [  {\bf f}_{ja} - m_a \nabla_j W_{aj} \right ]
\end{equation}
where, as noted earlier, the sum over the  constraint forces is zero. The rate of change of linear momentum of the fluid SPH particles is given by 
\begin{equation}
\sum_a m_a \frac{d {\bf v}_a}{dt} = \sum_a \sum_k \sum_{j \in S_k} m_a \left [{\bf f}_{aj} -  m_j\Pi_{aj} \nabla_a W_{aj}   \right ].
\end{equation}
noting that the sum over the pressure and viscous forces between fluid particles vanishes because of symmetry.

Recalling that 
\begin{equation}
\sum_a m_j {\bf f}_{ja} = - \sum_a m_a {\bf f}_{aj} \  {\rm and} \  \nabla_a W_{aj} = -\nabla_j W_{aj},
\end{equation}
we deduce that
\begin{equation}
\sum_k M_k \frac{d {\bf V}_k}{dt} + \sum_a m_a \frac{d {\bf v}_a}{dt}  = 0,
\end{equation}
which shows that the linear momentum
\begin{equation}
\sum_k M_k{ \bf V}_k + \sum_a m_a  {\bf v}_a,  
\end{equation}
is conserved. 

The angular momentum of the bodies is composed of the centre of mass angular momentum about some fixed origin, and the sum over each body of the spin angular momentum about the centre of mass  of body.  The time rate of change of the total centre of mass angular momentum is
\begin{equation}\label{eqn:bodangmom}
\sum_k M_k {\bf R}_k \times \frac{d {\bf V}_k}{dt} =  \sum_k {\bf R}_k \times \sum_{j\in S_k}  {\bf f}_j + \sum_k {\bf R}_k \times {\bf F}_k.
\end{equation}
The rate of change of the spin angular momentum is 
\begin{equation}\label{eqn:spinangmom}
\sum_k I_k \frac{ d \Omega}{dt} = \sum_k  \sum_{j\in S_k} ({\bf r}_j - {\bf R}_k ) \times {\bf f}_j + \sum_k \tau_k.
\end{equation}
The rate of change of the angular momentum of the fluid particles is 
\begin{equation}\label{eqn:fluidangmom}
\sum_a m_a  {\bf r}_a \times \frac{d {\bf v}_a}{dt} = \sum_a \sum_k \sum_{j \in S_k} m_a {\bf r}_a \times\left [  {\bf f}_{aj} -  m_j\Pi_{aj} \nabla_a W_{aj}  \right ],
\end{equation}
where, because of symmetry, the sum over pressure, and viscous  terms between fluid particles have vanished.  The rate of change of the total angular momentum (the sum of (\ref{eqn:bodangmom}), (\ref{eqn:spinangmom}) and (\ref{eqn:fluidangmom})) becomes
\begin{eqnarray}
\frac{dJ}{dt} &=&\sum_a \sum_k \sum_{j \in S_k} m_a({\bf r}_j - {\bf r}_a) \times{\bf f}_{aj} \nonumber \\
   &  & +\lambda_X^{1} \left(- Y_1  + \ell_1  \sin{(\theta_1)}  +Y_2 + \ell_2 \sin{(\theta_2)} \right) \nonumber\\
   & &+ \lambda_Y^{1} \left(X_1  - \ell_1  \cos{(\theta_1)}  - X_2 - \ell_2 \cos{(\theta_2)} \right)\nonumber \\
   & &+\lambda_X^{2} \left(- Y_2  +  \ell_2  \sin{(\theta_2)}  +Y_3 +  \ell_3 \sin{(\theta_3)} \right) \nonumber \\
   & & +\lambda_Y^{2} \left( X_2  - \ell_2  \cos{(\theta_2)}  +X_3 - \ell_3 \cos{(\theta_3)} \right). 
\end{eqnarray}
The first term vanishes because the boundary forces are radial and the  last four terms vanish because of the constraint conditions (\ref{eqn:xcons}) and (\ref{eqn:ycons}).
 
Finally we note that the previous arguments about conservation assume the time derivatives are exact. The actual conservation in the numerical simulations depends on the form of the time stepping algorithm. Linear momentum is always conserved to round off error, but the angular momentum conservation is less accurate because the Lagrange multipliers are calculated at the mid-point.  In our simulations we use periodic boundaries and these do not conserve angular momentum exactly. A detailed discussion of the conservation of angular momentum is given by Kajtar and Monaghan (2008).

\subsection{Remarks concerning external boundaries}

The SPH equations can be applied to the linked bodies moving in a channel, as is the case for many laboratory experiments on fish, or in a pond with an irregular boundary, by replacing the boundaries of the pond by boundary force particles as we have done for the rigid bodies. The SPH algorithm does not need to be changed if the linked bodies move through and out of a free surface, which would be required to mimic the motion of dolphins. This facility was used earlier for bodies hitting the water (Monaghan and Kos, 2000, Monaghan et al., 2003). In the present paper, where we compare our results with those of Kanso et al. (2005) and Eldredge (2008), we need to deal with an infinite medium. This cannot be done directly because it would require infinitely many particles. One alternative, and the simplest, is to replace fluids of infinite extent by periodic boundary conditions. These boundaries alter the solutions of the differential equations but the effects are small if the periodic cells are sufficiently large. We determine their effect by carrying out test calculations for successively larger domains.
\begin{figure}[htbp]
\begin{center}
\includegraphics[width= 0.6 \textwidth]{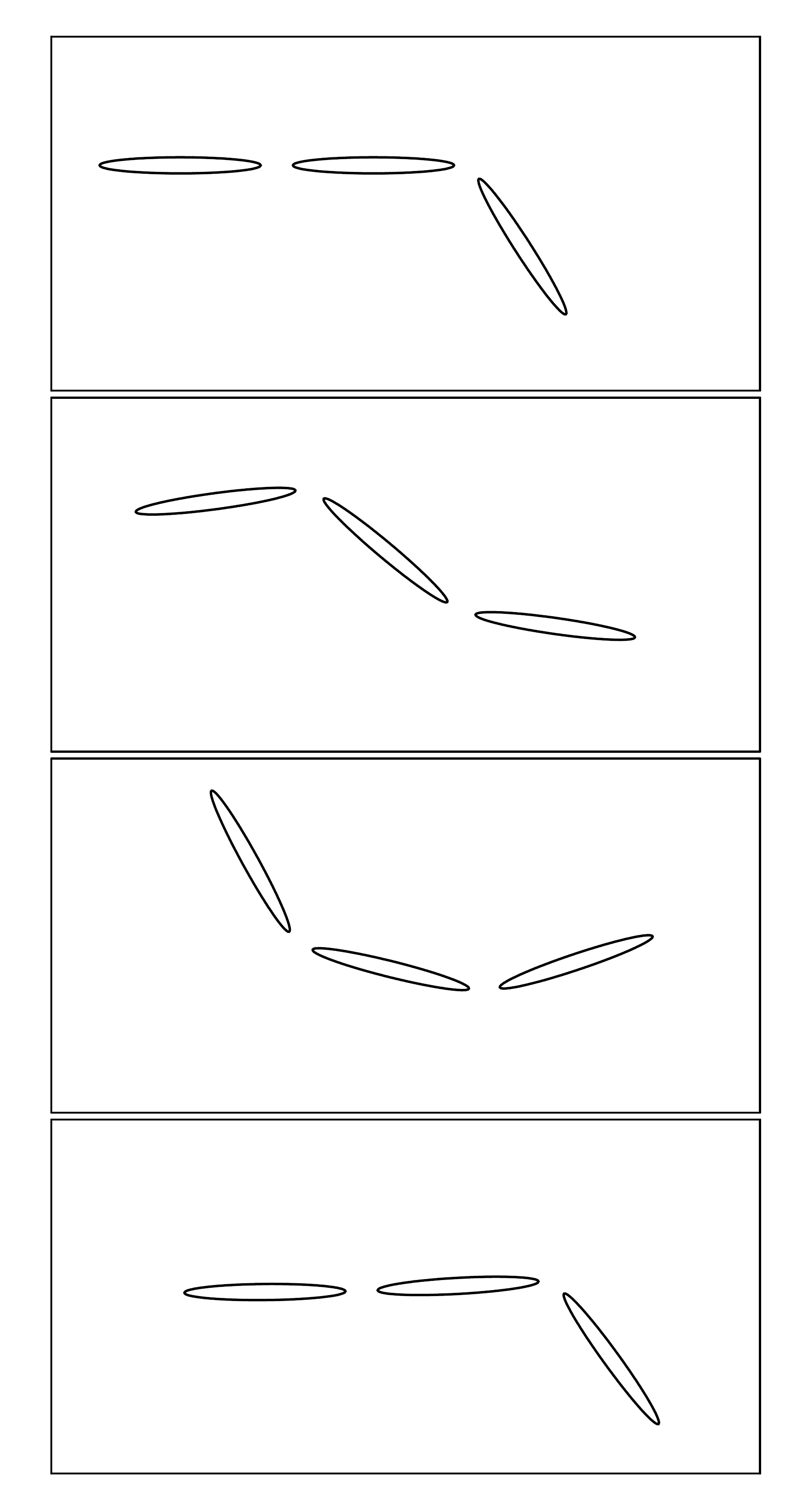}
\caption{The configuration of the bodies at time intervals separated by $2 \pi/3$ with time increasing from top to bottom.}
\label{fig:linkbodymotion}
\end{center}
\end{figure}
\section{The motion of the linked bodies}     
\setcounter{equation}{0}
We consider ellipses moving with the gait
\begin{eqnarray}
\varphi_1 &=&\theta_2(0) - \theta_1(0) +  \beta( \cos(\omega t) -1) \label{eqn:phi1}, \\
\varphi_2 &=&\theta_3(0) - \theta_2(0) + \beta \sin(\omega t) \label{eqn:phi2},
\end{eqnarray}
where throughout this section, we set $\beta=1$ and $\omega=1$. The ellipses have  semi-major axis $a=0.25$, semi-minor axis $b=0.1a$, and distance between the tip of the ellipse and the pivot $c=0.2a$. These dimensions, and the gait, are identical to those of Kanso et al. (2005) and Eldredge (2008) but we use a different notation for the angles.  The configuration of the ellipses is shown in Figure \ref{fig:linkbodymotion} at intervals of 1/3 of a period.

We define the Reynolds number $\Re$ by using the characteristic velocity $V = 2a\omega$ and the characteristic length scale $L=2a$, so that
\begin{equation}
\Re = \frac{4a^2\omega}{\nu}.
\end{equation}
The speed of sound $c_s = 20 a \omega$, and the boundaries of the ellipses were defined by boundary particles with spacing $ dp/4$ (Monaghan and Kajtar, 2009).  The motion takes place in a domain with periodic rectangular cells.

The motion of the linked bodies is characterised by the path followed by the centre of mass of the middle body. This path will be referred to as the `stride path'. The gait (\ref{eqn:phi1}) and (\ref{eqn:phi2}) is oscillatory  with period $P = 2\pi/\omega$ so that the stride path is oscillatory and has the shape of a zig zag.  We refer to the straight line distance between two consecutive lower points of this zig zag, travelled in time $P$, as a `stride length'.  The results to follow show that the stride length is,  in general, not constant, in agreement with the results of Eldredge.

 Kajtar and Monaghan (2008) showed that the SPH algorithm gave results in good agreement with experiments for a driven oscillating cylinder, and  for cylinders freely oscillating in a channel flow.  In this paper we describe three levels of further tests.  The first of these  is concerned with the convergence as the resolution is refined with a fixed periodic cell size, and convergence as the size of the cell is increased with fixed resolution ($\S 6.1 $ and $ \S6.2$).  The latter is to ensure that our comparison with the results of Kanso et al. (2005) and Eldredge (2008) is legitimate. The second level of tests is concerned with comparisons with the results of Eldredge by studying the stride length when the mass of the bodies is reduced and Kanso et al. by studying the stride length change as the viscosity coefficient is increased ($\S 6.3$ and $ \S 6.4$). The third level of tests shows that the numerical simulations agree with general scaling relations (\S 7).  

\subsection{Test of the convergence with resolution}

Throughout this section, we use a rectangular domain with periodic boundaries aligned with the $x$ and $y$ axes of a cartesian coordinate system. The ratio of the lengths of the sides of the domain, the aspect ratio, is  4:3. The fluid spans from $x_\mathrm{min} = 0$ to $x_\mathrm{max}$ along the horizontal axis, and from $y_\mathrm{min} = 0$ to $y_\mathrm{max}$ in the vertical axis. The initial coordinates of the centre of mass of the middle body were always $(X_2,Y_2) = (0.4x_\mathrm{max}, 0.6y_\mathrm{max})$.  For the SPH simulations the periodic boundaries were implemented by copying rows and columns of fluid particles $4h$ in width to the opposite boundary, top to bottom, left to right, and vice versa. This process guarantees that the fluid particles of interest in the rectangular domain get the correct rates of change in each time step. 

We ran the calculations for initial particle spacing $dp = 1/30$, 1/40, 1/50 and 1/60. The domain was of size $x_\mathrm{max} = 4$ and $y_\mathrm{max} = 3$. For these tests, the bodies were neutrally buoyant and $\Re=200$. The simulations for each resolution were run for the same time. 

\begin{figure}[htbp]
\begin{center}
\includegraphics[width=0.8\textwidth]{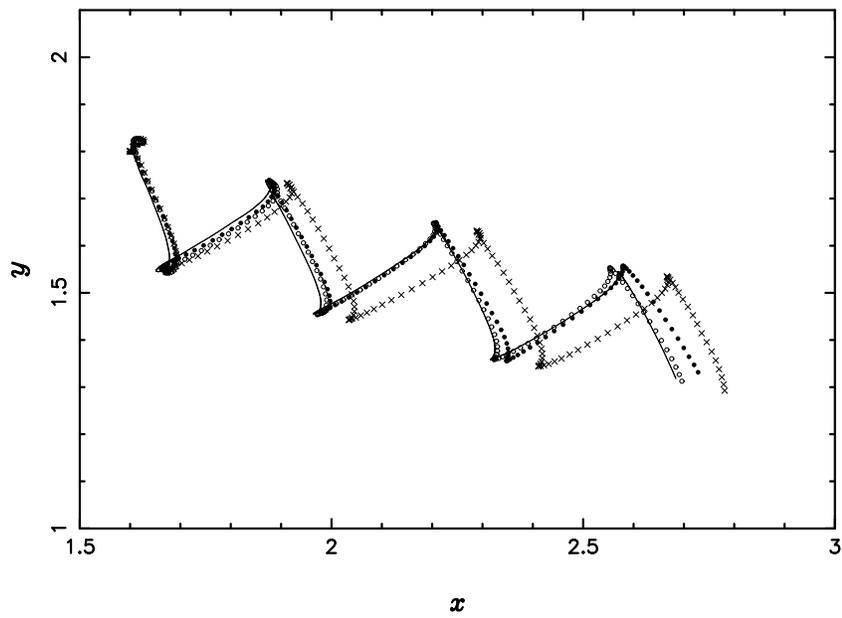}
\caption{The stride paths for different resolutions. The crosses denote the stride path with $dp = 1/30$. Open circles, filled circles and the solid line are for $dp = 1/40$, 1/50 and 1/60 respectively.}
\label{fig:respaths}
\end{center}
\end{figure}

The stride paths for the four values of $dp$ are plotted in Figure \ref{fig:respaths}. Note that the strides for the lowest resolution ($dp = 1/30$) are significantly longer than for the other three finer resolutions. The paths for $dp = 1/40$, 1/50 and 1/60 lie almost on top of one another although, because of the slight differences in average velocity, the differences increase with time and we note that the convergence is not monotonic i.e. the results for $dp=1/40$ are closer to those for $1/60$ than are those for $1/50$.  However, for the three smallest resolutions the relative difference between a stride length of one resolution and another is at most 5\% (for the third stride). Figure \ref{fig:respaths} also shows that the direction of the path is not sensitive to the resolution.  The results of this numerical test indicate that a fluid particle resolution of $dp = 1/40$ is sufficiently accurate to determine the stride path in length and direction to within 5\%.

\subsection{Test of convergence with domain size}

In order to determine a fluid domain size that adequately represents an infinite domain, the calculation was run for a number of periodic cell sizes with fixed $dp$. We ran the calculation with four different domain sizes with the same aspect ratio,  $x_\mathrm{max} \times y_\mathrm{max} = 4 \times 3$, $5 \times 3.75$, $6 \times 4.5$ and $7 \times 5.25$.  Again, the bodies were neutrally buoyant and $\Re=200$. The calculations were run for $dp = 1/40$, but we found that stride paths varied substantially from one case to the next. However, with a resolution of $dp = 1/60$, the stride paths show a smoother trend with increased domain size.

\begin{figure}[htbp]
\begin{center}
\includegraphics[width=0.8\textwidth]{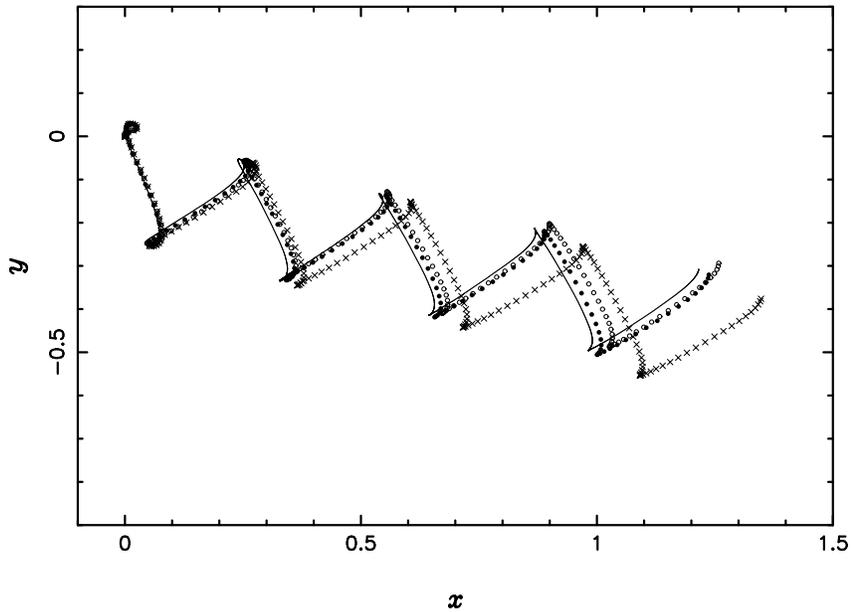}
\caption{The stride paths for different domain sizes with fixed $dp=1/60$. The crosses denote the stride path for domain size $x_\mathrm{max} \times y_\mathrm{max} = 4 \times 3$. Open circles, filled circles and the solid line are for $5 \times 3.75$, $6 \times 4.5$ and $7 \times 5.25$ respectively. Note that for the purposes of comparing the paths on this plot, the stride paths have been shifted so that they all begin at (0,0).}
\label{fig:domainpaths}
\end{center}
\end{figure}

The stride paths for the four domain sizes are plotted in Figure \ref{fig:domainpaths}. Note that the strides for the smallest domain are significantly longer than for the other three larger domains. These results indicate that a domain of size $5 \times 3.75$ is close to being sufficiently large for modelling an infinite domain. The distances travelled in three strides for the different domain sizes, and for the two resolutions are given in Table \ref{tab:domainstrides}. These values demonstrate the large variation for different domain sizes with $dp = 1/40$. For $dp = 1/60$, neglecting the smallest domain, the maximum relative difference is 3\% (between the $5 \times 3.75$ and $7 \times 5.25$ domains). For $dp = 1/40$ on the other hand, the maximum relative difference is 9\% (between the $5 \times 3.75$ and $6 \times 4.5$ domains).

\begin{table}[htdp]
\begin{center}
\begin{tabular}{|c|c|c|}
\hline
domain & $dp = 1/40$ & $dp=1/60$ \\
\hline
$4 \times 3$ & 0.9844 & 1.0031 \\
$5 \times 3.75$ & 0.9235 & 0.9214 \\
$6 \times 4.5$ & 0.8368 & 0.9149 \\
$7 \times 5.25$ & 0.8859 & 0.8932 \\
\hline
\end{tabular}
\end{center}
\caption{Distance travelled in three strides with different domain sizes, and for two different fluid resolutions.}
\label{tab:domainstrides}
\end{table}

Based on the numerical tests for the resolution and the domain size, we chose $dp = 1/60$ and a domain of size $6 \times 4.5$ for our subsequent production runs.
\subsection{Comparison with Eldredge}

Eldredge (2008) considers massless bodies, which are inconvenient to use with our algorithm (the expressions for the body velocities $V_k$ with $M_k=0$ are singular). We can, however, observe the trend in the motion of the linked bodies as their mass $M \to 0$. For neutrally buoyant elliptical bodies, the masses are  $M_0 = \rho_0 \pi ab$, and moment of inertia $I = M_0 (a^2 + b^2)/4$. We ran the simulation for a number of body masses in the range $0.5M_0 \leq M \leq 5.0M_0$ in order to determine a relationship between the mass and the stride length. For these simulations $dp=1/60$. The Reynolds number $\Re = 200$ is the same as in the calculation of Eldredge.

\begin{figure}[htbp]
\begin{center}
\includegraphics[width=0.8\textwidth]{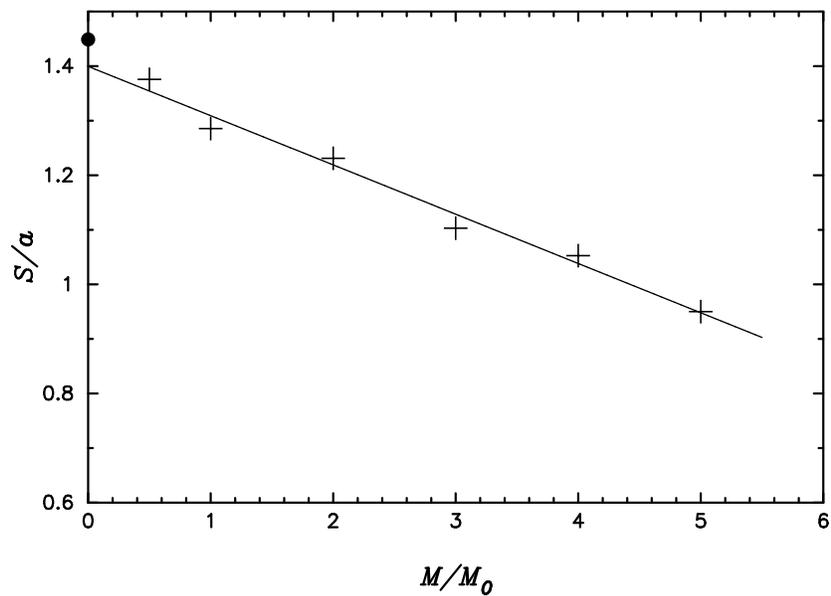}
\caption{Stride lengths $S$, scaled with the length parameter $a$, for the motion in fluid with constant viscosity and fixed gait but  a range of body masses. The crosses denote the SPH results and the filled circle denotes the result of Eldredge. The line of best-fit is also shown.}
\label{fig:eldredge-comp}
\end{center}
\end{figure}

Eldredge reports that the massless linked bodies have a stride length of $1.45a$. The stride lengths from the SPH simulations, as well as the result of Eldredge are plotted in Figure \ref{fig:eldredge-comp}. The line of best-fit  shows that there is a linear trend toward Eldredge's $M=0$ result. 

The results of Eldredge show that the stride lengths vary from stride to stride. The second stride is longer than the first, and the third is longer than the second. Our results show a similar behaviour.  We find that the second stride length is larger than the first typically by $\sim 18-28\%$, and the third stride length is larger than the second by $\sim 5-16\%$. The equivalent results of Eldredge are  $20\%$, and $10\%$ which is similar to our results.  For the inviscid case (discussed below),  Kanso et al. show that the stride length is constant.

Eldredge estimated the stride length by taking the average of the second and third strides, and we followed the same procedure in generating the results of Figure \ref{fig:eldredge-comp}.

\begin{figure}[htbp]
\begin{center}
\includegraphics[width=1.0\textwidth]{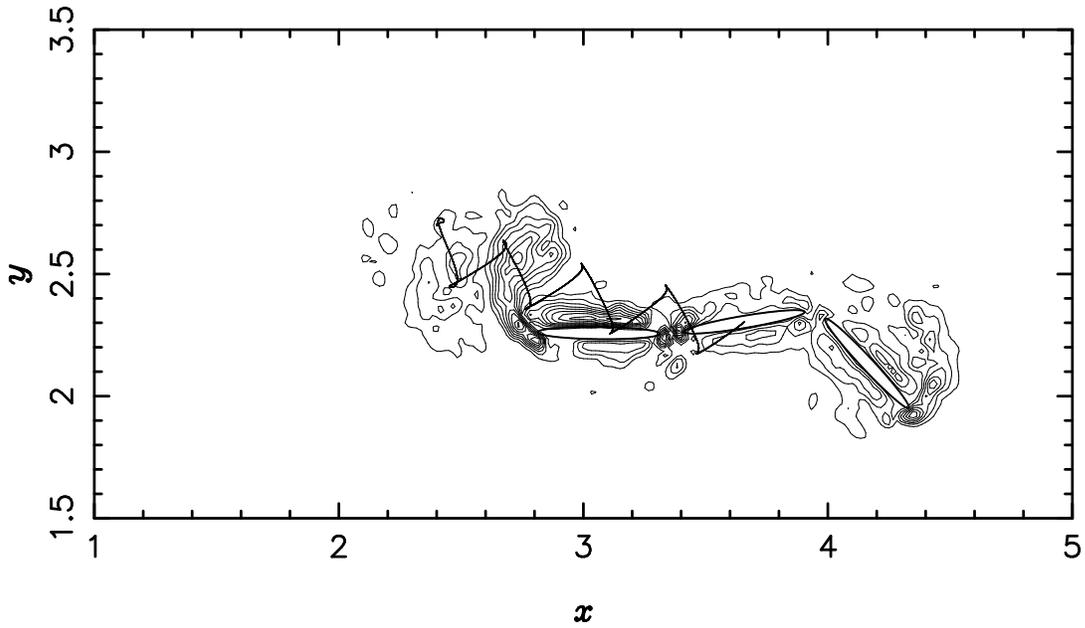}
\caption{Vorticity field generated by the swimming linked bodies with $M = 0.5M_0$. This plot is at time $t \sim 25$. The vorticity contours have values in the range -5 to 5 with 40 levels. The stride path is also shown.}
\label{fig:viscousvort}
\end{center}
\end{figure}

The vorticity generated by the motion of the linked bodies after time $t \sim 25$ is shown in Figure \ref{fig:viscousvort}. This plot can be compared to the last frame of Figure 10 of Eldredge (2008). Note however, that for this figure $M = 0.5M_0$, whereas Eldredge has massless bodies. The contours of Eldredge are much smoother than those shown in Figure \ref{fig:viscousvort}, but the main features are recognisable. There is one large, and one smaller eddy near the rear of the linked bodies, which are in the same positions as with Eldredge, and there is an intense eddy generated by the front body. The stride path is in good agreement with Eldredge.

Finally, we note that the vortex particle spacing in  Eldredge's simulation is typically $a/50$ compared to our $a/15$. Eldredge has 280 panels on each body which is close to the 244 boundary particles per body in the SPH calculation.


\subsection{Comparison with Kanso}

Kanso et al. (2005) consider the motion of three linked ellipses in an inviscid fluid. Although the SPH algorithm is only stable with non-zero viscosity we can study the stride length variation with change in the viscosity and estimate the stride length for zero kinematic viscosity coefficient as $\nu \to 0$. We ran the simulation with neutrally buoyant bodies for viscosity in the range $5 \times 10^{-5} \leq \nu \leq 5 \times 10^{-3}$, which corresponds to a range in Reynolds number of $50 \leq \Re \leq 5000$. The calculations were run with neutrally buoyant bodies.

\begin{figure}[htbp]
\begin{center}
\includegraphics[width=0.8\textwidth]{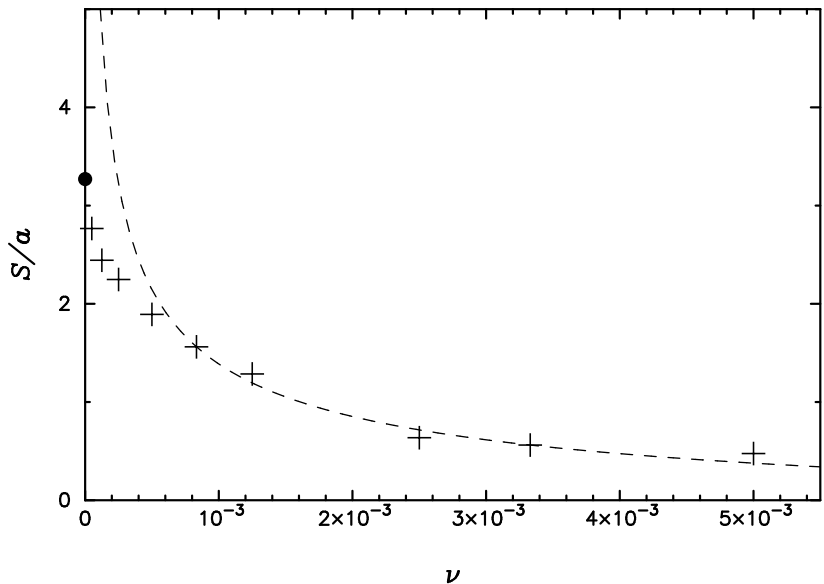}
\caption{Stride lengths, in terms of length parameter $a$, for the motion in a fluid with different viscosities $\nu$ but constant gait and mass. The crosses denote the SPH results and the filled circle denotes the result of Kanso et al. The dashed curve is based on an analytical result by Taylor, which is not expected to be valid for very high Reynolds number. Note that if $\nu = 10^{-3}$ then $
Re = 250.$}
\label{fig:kanso-comp}
\end{center}
\end{figure}

Kanso et al. find that the stride length for the neutrally buoyant bodies is $3.27a$. This result, as well as the stride lengths from the SPH simulations, are plotted in Figure \ref{fig:kanso-comp}. The SPH results show a trend toward the $\nu = 0$ case of Kanso. In some respects the agreement is remarkable because there are significant differences between the inviscid and non-inviscid cases. For example the flow produced by an oscillating cylinder changes dramatically as the Reynolds number changes from  small $\Re \sim 10$ to large $\Re \sim 1000$ though the time averaged drag terms are nearly constant for $100 < \Re <10000$.  In the inviscid case the fluid motion produced by a system of oscillating linked bodies will cease the instant they stop oscillating, while in the viscous case, the motion will continue though it will be damped.  And as discussed in the previous section, the strides increase in length both for our calculations and those of Eldredge whereas those of Kanso et al. are constant. These results suggest that when $\Re \gtrsim 1000$ the average motion of the linked bodies is determined primarily by added mass effects as discussed by Saffman (1967) for swimming by shape change in an inviscid fluid.

The stride lengths plotted in Figure \ref{fig:kanso-comp} were calculated as described in the last section. In addition to the SPH calculations we have plotted in Figure \ref{fig:kanso-comp} an estimate of the variation of the stride length with viscosity based on an analytical result obtained by Taylor (1952) in his discussion of the swimming of long slender bodies. A curve was fitted for the stride length of the form
\begin{equation}\label{eqn:kanso-fit}
S = \frac{s_1}{\nu^{1/2}} + s_2,
\end{equation}
where $s_1 $ and $s_2$ are arbitrary constants (since we have fixed $\beta$) determined by fitting to our data set. The form of (\ref{eqn:kanso-fit}) is determined from (\ref{eqn:taylorv-general}), which will be discussed in the next section. For the present case we determine the coefficients using two points from the SPH results. We find $s_1 = 0.05774$ and $s_2 = -0.4386$. The curve shows good agreement in the higher viscosity range where $\nu \gtrsim 5 \times 10^{-4}$ and $\Re \le 500$. We do not expect (\ref{eqn:kanso-fit}) to be valid for very high Reynolds number.

\section{Speed and power scaling relations}

The speed with which the linked bodies move through the fluid depends upon a number of parameters. As already seen, the speed depends (at least) upon the mass of the bodies and the fluid viscosity. Additionally, we expect the speed to depend upon the ratios $a/b$, $a/c$, the frequency $\omega$, and the amplitude $\beta$. Similarly, we expect the power expended to be dependent on these parameters.  Because the fluid is treated as slightly compressible there is a further non dimensional quantity $\omega a /c_s  $ typically equal to 1/20 in our calculations.  We neglect  contributions from this quantity.

The speed of the linked bodies, $V$, was estimated from the stride length divided by the time taken to complete the stride. Following the approach from the previous section, we take the stride length to be the average of the second and third strides. The time taken to complete a stride is  $2\pi/\omega$.

In a biological creature the power expended for locomotion is provided by the actions of the muscles which themselves depend on their body chemistry. In the case of a marine robotic vehicle  the energy is provided by the engines within the vehicle.  In our formulations the energy can be estimated from the constraint forces in the equations of motion. We calculate the average power $\mathcal P$ expended by the linked bodies over the time interval $t_1$ to $t_2$ from the expression

\begin{equation} \label{eqn:powerdef}
{\mathcal P} = \frac{1}{t_2-t_1} \int _{t_1}^{t_2}\left(\sum_k \mathbf{V}_k \cdot \mathbf{F}_k + \sum_k \Omega_k \tau_k \right) dt,
\end{equation}
where $\mathbf{F}_k$ and $\tau_k$ are the constraint forces and torques on body $k$ respectively. Substituting the constraint forces and torques (\ref{eqn:cforce1}-\ref{eqn:ctorque3}) into (\ref{eqn:powerdef}) gives
\begin{equation} \label{eqn:powercons}
{\mathcal P} = \frac{1}{t_2 - t_1} \int _{t_1}^{t_2}\left( \lambda_\theta^{(1)} ( \Omega_2 - \Omega_1 ) + \lambda_\theta^{(2)} ( \Omega_3 - \Omega_2 ) \right) dt.
\end{equation}

 $\mathcal P$ was calculated by numerical integration over the time of interest (in this case, the time taken to complete the first three strides).

 To simplify the velocity and power relations, we study the motion with bodies of fixed mass, fixed lengths $a$, $b$, and $c$ and a fixed periodic-domain size. We then expect the speed to be given by an expression of the following form
\begin{equation}\label{eqn:velrelation}
V = \omega a F(\Re, \beta).
\end{equation}
\begin{figure}[htbp]
\begin{center}
\includegraphics[width=0.7\textwidth]{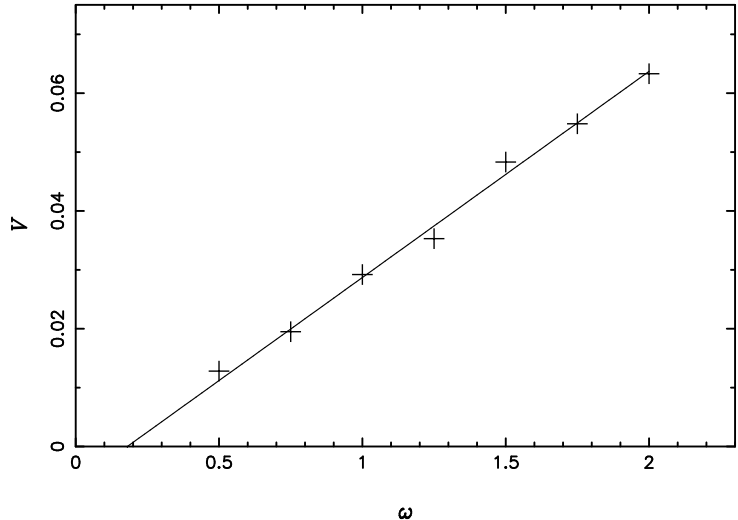}
\caption{Velocity of the linked bodies for different frequencies with constant Reynolds number and amplitude. The SPH results are shown by the crosses and the continuous curve is a best fit straight line.}
\label{fig:constreynum-vel}
\end{center}
\end{figure}
Since the work done by the constraints is proportional to $\omega^2 a^2$, and $t \propto 1/\omega$, we expect a power relation of the following form
\begin{equation}\label{eqn:powrelation}
{\mathcal P} = \beta^2 \omega^3 a^2 G(\Re, \beta).
\end{equation}

It is useful to compare these scaling relations with the analysis of Taylor (1952).  He shows that  an infinite, flexible cylinder, along which a wave of amplitude $B$ and wavelength $\lambda$ propagates with speed $U= \lambda  \omega/(2 \pi)$, moves with an average forward velocity $V$ in a viscous fluid given by  
\begin{equation}\label{eqn:taylorexp}
C_D R_1^{1/2} C(\alpha) = \frac{ 5.4 \gamma(\alpha)I_1(\alpha)}{(1-n)^{3/2}} - \frac{1}{(1-n)^{1/2} } ( 5.4 I_2(\alpha) + 4 I_3(\alpha)),
\end{equation}
where $n = V/U$, $C_D$ is a drag coefficient,  and the functions $\gamma$, $C$, $I_1$, $I_2$, and $I_3$ are given by integrals. The quantity $\alpha$ is given by $\tan{\alpha} =2 \pi B/\lambda$ and when $\alpha$ is small enough $\alpha \sim 2 \pi B/\lambda$. For our oscillating bodies  $\lambda \sim 6a$, so that $\alpha \sim B/ a$.  Because $B/a$ is close to our amplitude $\beta$ we replace $\alpha $ by $  \beta$ to convert Taylor's formula to a form appropriate for our system. $R_1 = Ud/\nu$ is a Reynolds number where the characteristic length is $d$ the diameter of the cylinder. Taylor's formula is an example of the relation (\ref{eqn:velrelation}) with $a$ replaced by $\lambda$.
\begin{figure}[htbp]
\begin{center}
\includegraphics[width=0.7\textwidth]{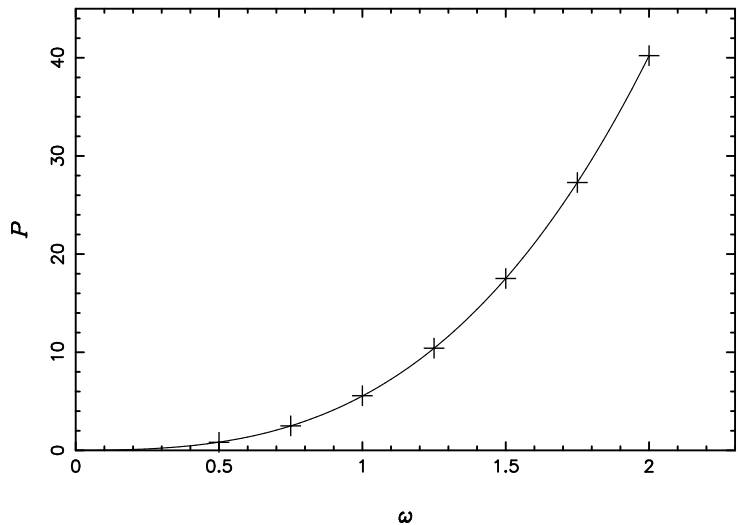}
\caption{Power expended by the linked bodies for different frequencies with constant Reynolds number and amplitude. The SPH results are shown by crosses and the continuous curve is a best fit cubic.}
\label{fig:constreynum-pow}
\end{center}
\end{figure}
Our oscillating ellipses are similar to a small section of Taylor's oscillating cylinder and this suggests that Taylor's formula might provide a useful model for the scaling relations appropriate to the linked ellipses even though his calculations are for motion in three dimensions and ours are for motion in two dimensions. To that end we replace $R_1$ by our Reynolds number $\Re$ and expand  Taylor's formula assuming $n$ is small.  We then find, with $U  \propto \omega a$, that
\begin{equation}\label{eqn:velrelationb}
\frac{2 \pi V}{\omega a} = A(\alpha) \Re^{1/2} + B(\alpha)
\end{equation}
where the left hand side is the stride length scaled in units of $a$ and a constant of proportionality has been absorbed into $A$ and $B$.  In the following we replace $\alpha$  by $\beta$.  The continuous curve in Figure \ref{fig:kanso-comp}, which applies to the case of constant $\beta$ and $\omega$, was obtained by fitting the parameters $A$ and $B$ using two values of the viscosity.  It can be seen that this gives a good fit to our results except at the lowest values of the viscosity coefficient.  Furthermore, for constant viscosity and amplitude, the variation of $V$ with frequency deduced from (\ref{eqn:velrelationb}) is 
\begin{equation}
V = c_1 \omega ( \omega^{1/2} + c_2),
\end{equation}
where $c_1$ and $c_2$ are constant for fixed amplitude, viscosity and length $a$.

If $\alpha \ll 1$, the integrals in Taylor's formula can be expanded in a power series in $\alpha$.  If this is done, and we replace $\alpha$ by $\beta$, we find that 
\begin{equation}\label{eqn:taylorv-general}
\frac{V}{\omega a} = k_1 \Re^{1/2} \beta^{5/2} + k_2\beta^2,
\end{equation} 
where $k_1$ and $k_2$ are constants. These formula, with suitable values for the constants, give a very satisfactory fit to our results.

Taylor (1952) also derives an expression for the power generated.  If $n \ll1$, and the amplitude $\beta<1$ we can expand Taylor's formula to find (we replace his notation W for the power by $\mathcal P$) that 
\begin{equation}
{\mathcal P} = \frac{\omega^3 a^3}{\Re^{1/2}} (c_1' \Re^{1/2} \beta^2 + c_2' \beta^{5/2}).
\end{equation}
When the viscosity and the lengths $a$, $b$ and $c$ are  constant we can write this as
\begin{equation}\label{eqn:taylorp-general}
{\mathcal P} = c_1''\omega^{3} \beta^2 + c_2'' \omega^{5/2} \beta^{5/2}.
\end{equation}
Taylor's expressions for the velocity and power agree, as expected, with the general scaling relations (\ref{eqn:velrelation}) and (\ref{eqn:powrelation}).
\subsection{Speed and power for constant Reynolds number}
From the scaling relations  we expect for a fixed amplitude and Reynolds number that $V \propto \omega$ and $\mathcal{P} \propto \omega^3$. In order to test these relations  the mass of each ellipse was kept constant (neutrally buoyant), with constant lengths $a=0.25$, $b=0.2a$ and $c=0.2a$ and Reynolds number 200, with $\beta=1$. The frequencies were in the range $0.5 \leq \omega \leq 2.0$. In order to keep the Reynolds number fixed, the fluid viscosity was changed with the frequency of oscillation according to $\nu = 1.25\times 10^{-3}\omega$. Additionally, the speed of sound required for the equation of state was  constant with the value $c_\mathrm{s} = 10 \times 2a \times \omega_\mathrm{max} = 10$ where $\omega_\mathrm{max} = 2$ is the maximum frequency used.

Figure \ref{fig:constreynum-vel} shows that there is an approximate linear relationship between the velocity and the frequency, which is in substantial agreement with (\ref{eqn:velrelation}) except that the velocity vanishes for $\omega \simeq 0.2$. Figure \ref{fig:constreynum-pow} shows the  power $\mathcal {P}$ against $\omega$. The continuous curve is a cubic polynomial in agreement with (\ref{eqn:powrelation}). 


\subsection{Speed and power for constant viscosity}
For the case where the viscosity is fixed  we can be guided by Taylor's formula. We ran the calculations, again with neutrally buoyant bodies, and the same body length parameters. We chose $\nu = 1.25 \times 10^{-3}$, such that $\Re = 200\omega$. We ran the calculations for a number of frequencies and amplitudes in the range $0.5 \leq \omega \leq 2.0$ and $0.3 \leq \beta \leq 1.6$. Note that for $\beta > \pi/2$, the angle between two consecutive bodies is less than $90^\circ$. This case has no physical analogue to the long slender body of Taylor. 

\begin{figure}[htbp]
\begin{center}
\includegraphics[width=0.7\textwidth]{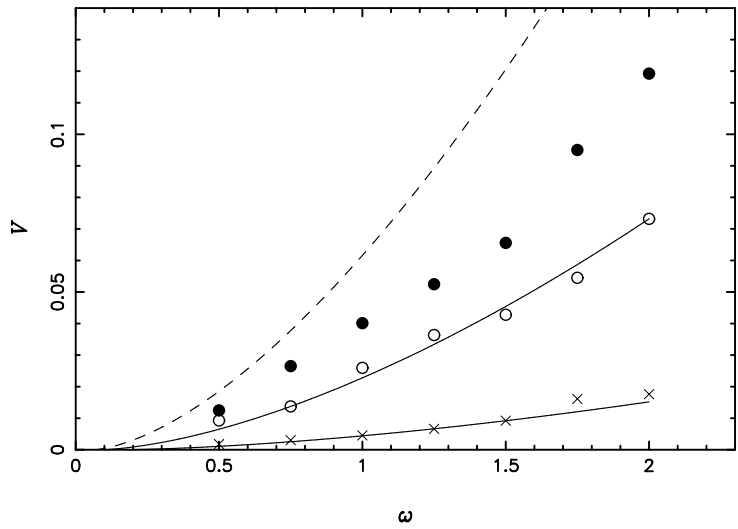}
\caption{Velocity curves of constant viscosity and amplitude against frequency. Crosses, open circles and filled circles are for $\beta$ = 0.5, 0.9 and 1.3 respectively. The curves are from (\ref{eqn:taylorv-general}) with constants fitted from the set with $\beta = 0.9$. The dashed curve is for $\beta = 1.3,$ which is outside the range for which (\ref{eqn:taylorv-general}) is valid.}
\label{fig:vel-freq}
\end{center}
\end{figure}

\begin{figure}[htbp]
\begin{center}
\includegraphics[width=0.7\textwidth]{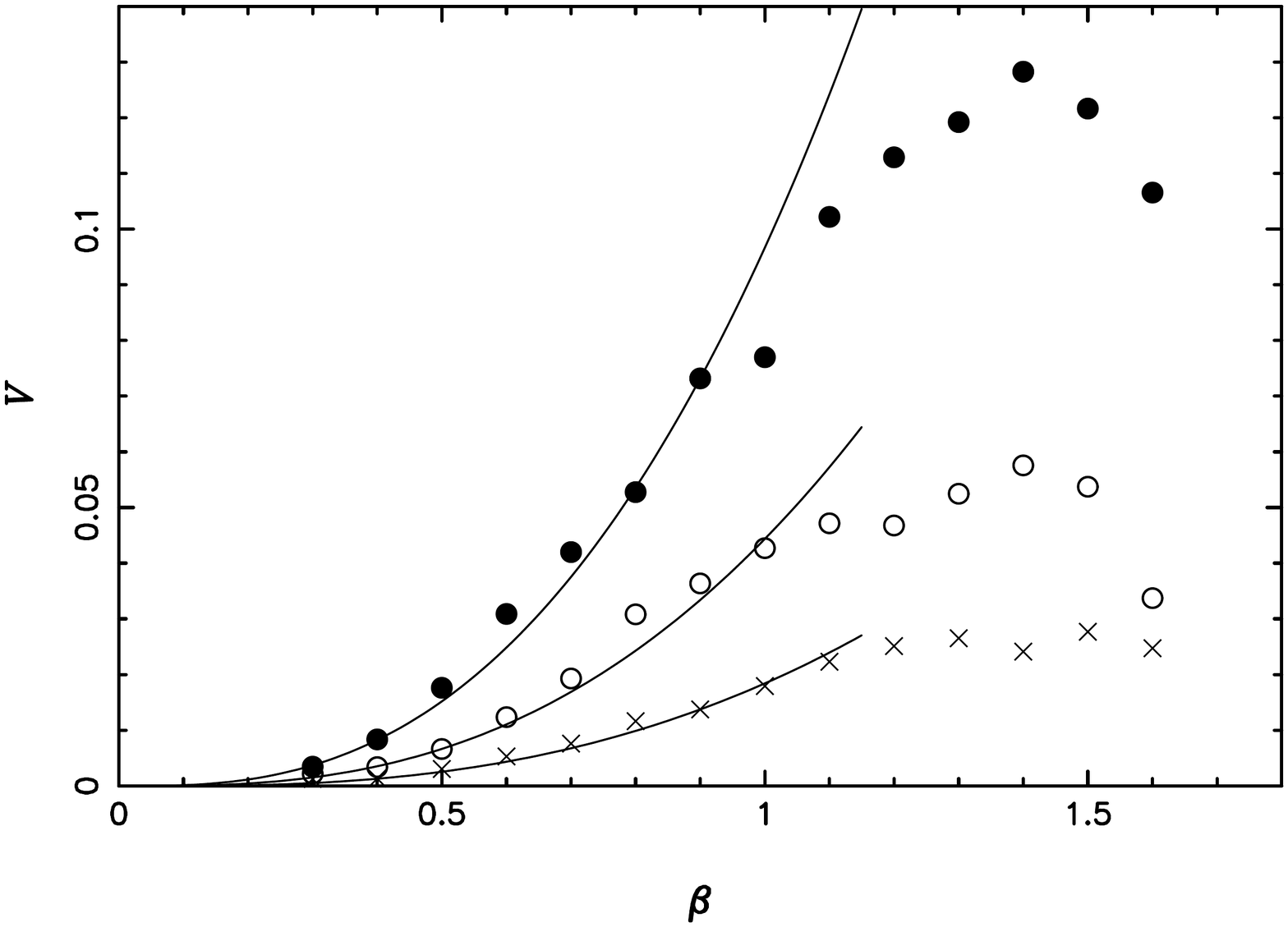}
\caption{Velocity curves of constant viscosity and frequency against amplitude. Crosses, open circles and filled circles are for $\omega$ = 0.75, 1.25 and 2.0 respectively. The curves are from (\ref{eqn:taylorv-general}), which is appropriate for $\beta < 1$.}
\label{fig:vel-amp}
\end{center}
\end{figure}

The speed is plotted against $\omega$ for three amplitudes 0.5, 0.9, and 1.3 in Figure \ref{fig:vel-freq}. We used two data points from the $\beta=0.9$ data set in order to determine the constants $k_1$ and $k_2$ in (\ref{eqn:taylorv-general}) which can then be written.

\begin{equation}
V = 0.043396 \omega^{3/2} \beta^{5/2} -0.013039 \omega \beta^2.
\end{equation}
As Figure \ref{fig:vel-freq} shows, this gives a good fit to the $\beta = 0.5$ and 0.9 data. The (dashed) curve for $\beta=1.3$ does not agree with the SPH results, but we do not necessarily expect it to, since (\ref{eqn:taylorv-general}) is only valid for $\beta < 1$. The velocity is plotted against $\beta$ for fixed frequencies in Figure \ref{fig:vel-amp}. The curves on these plots are again from (\ref{eqn:taylorv-general}), using the same values for $k_1$ and $k_2$ as determined previously. Once again we see that the curves give a good fit for $\beta < 1$. We found that the peak velocity is achieved with $\beta \sim 1.4$ for all of these cases. The velocity appears to be smaller when the bodies swing through $90^\circ$, or more, relative to one another. 

\begin{figure}[htbp]
\begin{center}
\includegraphics[width=0.7\textwidth]{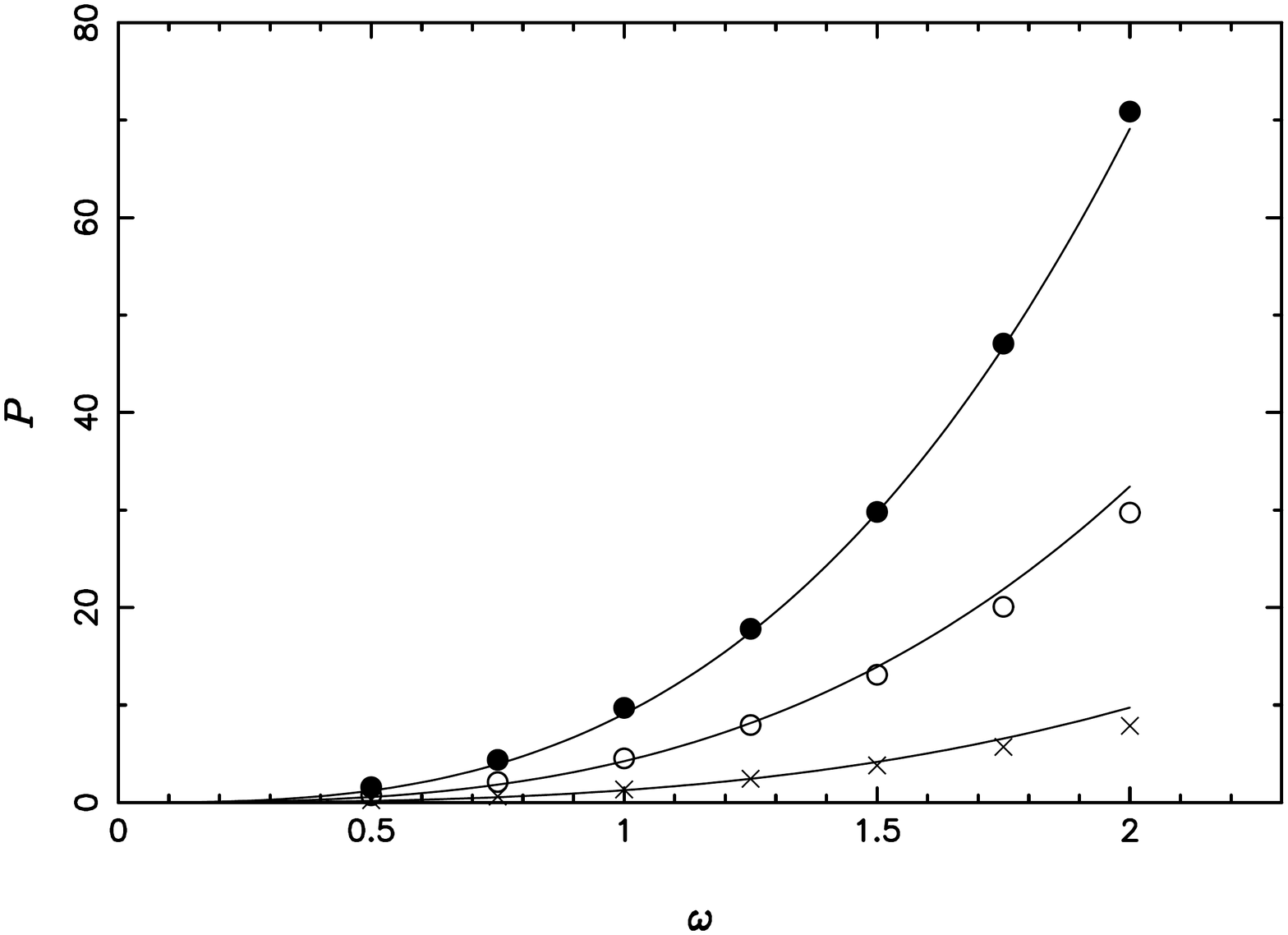}
\caption{Average power curves of constant viscosity and amplitude against frequency. Crosses, open circles and filled circles are for $\beta$ = 0.5, 0.9 and 1.3 respectively. The curves are from (\ref{eqn:taylorp-general}).}
\label{fig:pow-freq}
\end{center}
\end{figure}

\begin{figure}[htbp]
\begin{center}
\includegraphics[width=0.7\textwidth]{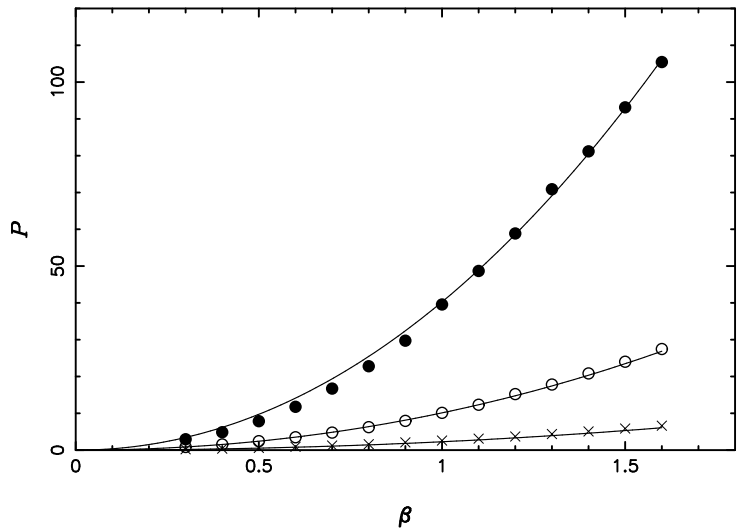}
\caption{Average power curves of constant viscosity and frequency against amplitude. Crosses, open circles and filled circles are for $\omega$ = 0.75, 1.25 and 2.0 respectively. The curves are from (\ref{eqn:taylorp-general}).}
\label{fig:pow-amp}
\end{center}
\end{figure}

The average power is plotted against $\omega$ for fixed amplitudes in Figure \ref{fig:pow-freq}. As with the velocity, we chose two data points from the $\beta= 0.9$ data set in order to determine the constants $c_1'' = 4.469$ and $c_2'' =0.798$ in (\ref{eqn:taylorp-general}).  This gives a very good fit to the SPH results. The power is plotted against $\beta$ for fixed frequencies in Figure \ref{fig:pow-amp}. Again, (\ref{eqn:taylorp-general}) with the same values for $c_1''$ and $c_2''$ gives a very good fit to the SPH results.

\subsection{An application to the swimming of a leech}
\begin{figure}[htbp!]
\begin{center}
\includegraphics[width= 0.4 \textwidth]{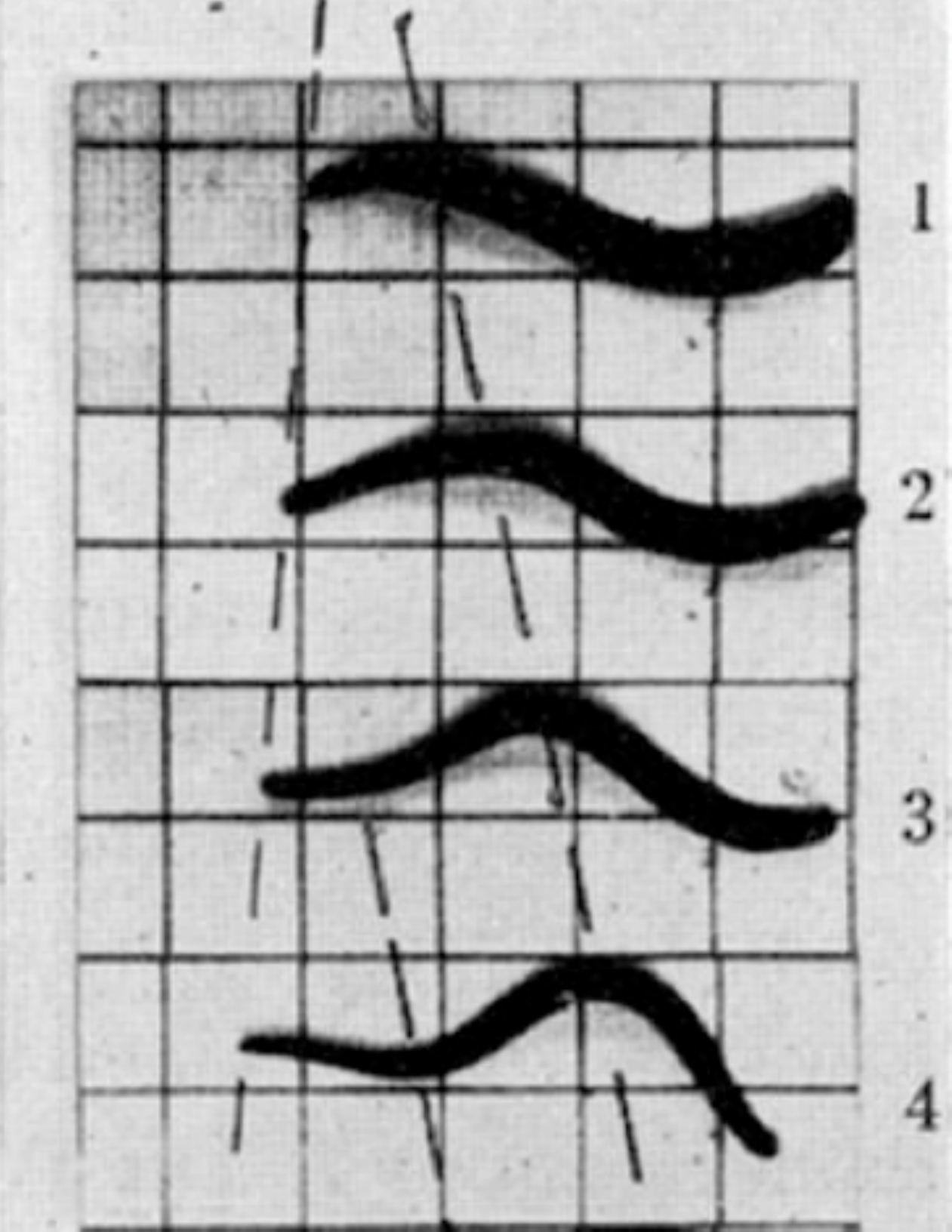}
\caption{Frames showing the motion of a leech against a background of squares of side 2 cm.  The time interval between frames is 15 s.  The image is part of  Figure 7 in Taylor's paper. The shape of the leech is roughly similar to that of our three oscillating ellipses.}
\label{fig:taylorleech}
\end{center}
\end{figure}

Taylor applied his formula to estimate the speed of a leech which swims with shape changes shown in Figure \ref{fig:taylorleech}. These shape changes only roughly approximate a sine wave down a very long cylinder since Taylor does not include end effects and, in any case the motion is only approximately sinusoidal (see for example frame 4).  We can estimate the speed by making appropriate adjustments to our scaling formulae. Since these mimic Taylor's the only issue is whether the constants calculated from our simulations allow us  to calculate the speed for an animal with different length scales.  

 Returning to the relation (7.12), and determining the constants by fitting to our results for  we find
\begin{equation}
V = 0.01227\left ( \frac{2a^2}{\sqrt{ \nu}} \right ) \omega^{3/2} \beta^{5/2} - 0.05216 \omega a \beta^2.
\label{eqn:leechvel}
\end{equation}

The leech Taylor considered was approximately 0.08 m long, and travelled with a velocity 0.043 m/s. The gait of the leech produces a wave like motion along its body with an average speed  0.153 m/s which we estimate as being equivalent to 
\begin{equation}
U = \frac{\omega}{2 \pi} \lambda \sim \omega a,
\label{eqn:U}
\end{equation}
where we take the wavelength as the length of the leech and set this to be approximately $6a$, the total length of our ellipses.  This gives $\omega \sim 12$ $\mathrm{s}^{-1}$. Taylor finds that the average value of $B/\lambda  = 0.089$ so that $\alpha = 0.56$. Consistent with out earlier discussion, we can take $\beta = 0.56$.  The thickness of the leech changes as it moves with an average diameter estimated by Taylor to be 0.0055m so that, for our ellipses, we can estimate $b = 0.0027$. The estimate of $a$ for the leech is 0.08/6 and the ratio $a/b \sim 5$ as in our simulations. The viscosity coefficient is  $10^{-6}$ $\mathrm{m}^2/\mathrm{s}$.  Substituting these values into (\ref{eqn:leechvel})  with $\beta = 0.56$, we find $V = 0.037$ m/s, which compares favourably with Taylor's estimate from the experiment of 0.043 m/s. It might be thought that this agreement is a lucky coincidence but, taken with the agreement of our results with formulae modelled on Taylor's expression,  it does suggest that the speed of  a three dimensional long thin body is similar in form to that of three linked, long, thin ellipses in two dimensions. 


\section{Optimal motion with constant viscosity}
\setcounter{equation}{0}

\begin{figure}[htbp]
\begin{center}
\includegraphics[width=1.0\textwidth]{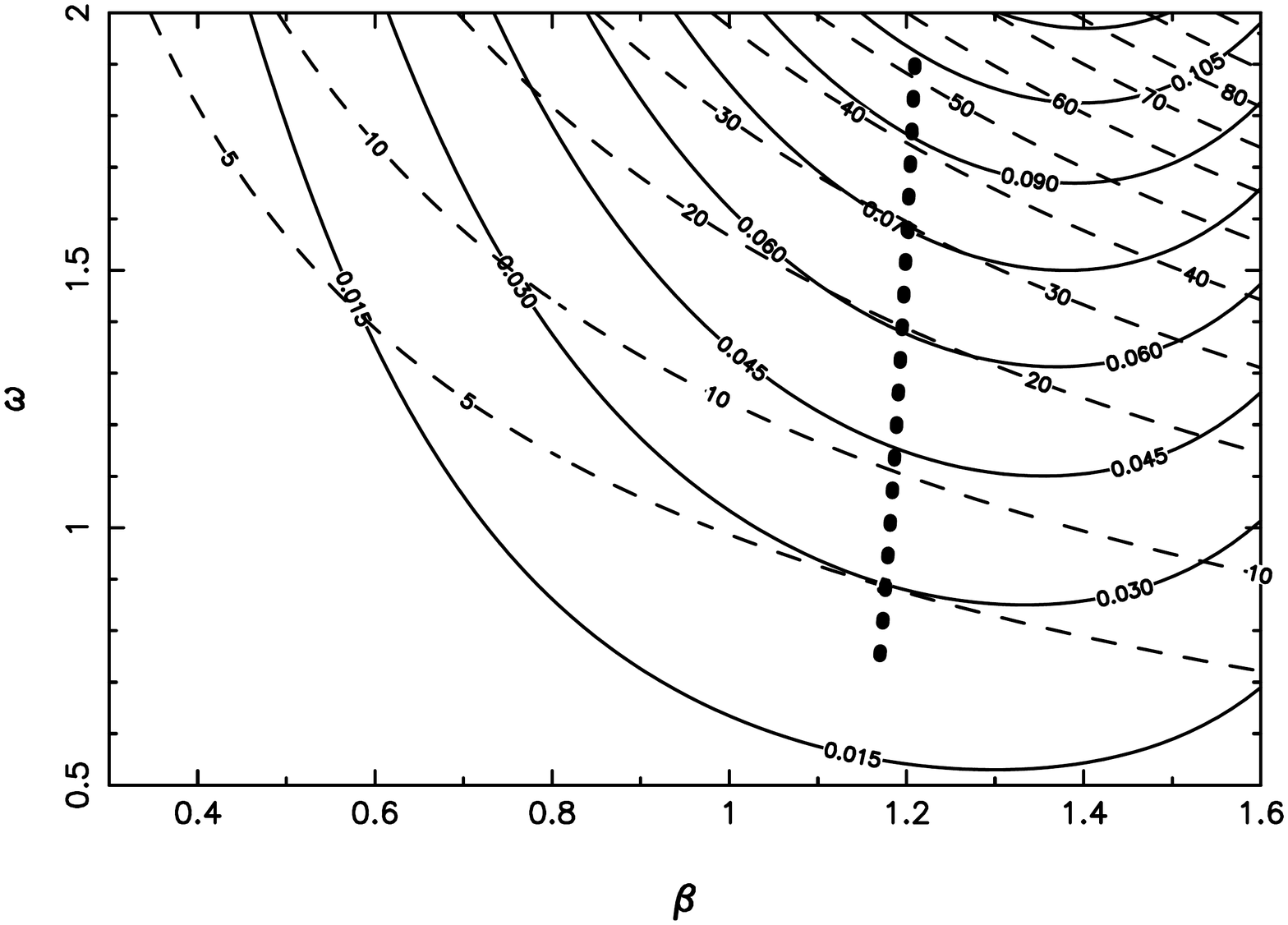}
\caption{Velocity and average power contours on a frequency-amplitude plot. The solid lines are the velocity contours, and the dashed lines are the power contours. The thick dotted line is the curve of optimal motion.}
\label{fig:optmotioncontours}
\end{center}
\end{figure}

We now ask the question: for a given fluid viscosity, body size and body mass configuration, what is the frequency and amplitude to move with a given speed, while expending the least power? We do not attempt to obtain the optimum frequency and amplitude for all possible ellipses. Instead we have the more modest aim of determining if an optimum set of parameters exists for a typical set of ellipses.  With this in mind we consider neutrally buoyant bodies, with $a=0.25$, $b=0.2a$ and $c=0.2a$. The kinematic viscosity is $\nu = 1.25 \times 10^{-3}$, so that $\Re = 200\omega$.  In Figure \ref{fig:optmotioncontours} we show the contours of constant velocity and constant power in the $(\omega, \beta)$ plane.  These contours were obtained by using the results of the simulations to fit the velocity with polynomials of the form
\begin{equation}
  V = \sum_{i=1}^2 \sum_{j=1}^4  C_{ij}  \omega^i  \beta^j,
\end{equation}
and the power $\mathcal{P}$ with the function $c \beta^2 \omega^3$.  While (8.1) does not include the fractional powers we derived by comparison with Taylor's formula it is still possible to a satisfactory over the whole range of $\omega$ and $\beta$. It is clear from Figure \ref{fig:optmotioncontours} that there is a set of values of $\omega$ and $\beta$ which will give a specified speed with minimum power.  The dashed line in Figure \ref{fig:optmotioncontours} gives the optimum set of $\omega$ and $\beta$.  This line was calculated by traversing a contour of constant power and finding the $\omega$ and $\beta$ which give the maximum speed. It is interesting to note that the optimal motion is close to $\beta \simeq 1.2$ regardless of the frequency.

We can estimate some properties of the contours in Figure \ref{fig:optmotioncontours} without detailed numerical calculations. For constant viscosity,  we can estimate from (\ref{eqn:taylorv-general}), for fixed lengths $a$, $b$ and $c$ that 
\begin{equation}
\beta^2 \propto \frac{V}{\omega ( \omega ^{1/2} + k_1')},
\end{equation}
where we have replaced $\beta^{5/2}$ by $\beta^2$ which is reasonably accurate for $0.5 < \beta <1$, and $k_1'$ is a constant.  Similarly from (\ref{eqn:taylorp-general}) we estimate
\begin{equation}
\beta^2 \propto \frac{\mathcal{P}}{ \omega^{5/2}(\omega ^{1/2}+ k_2')}.
\end{equation}
Where $k_2'$ is a constant. These expressions show that $\beta$ increases faster as $\omega$ decreases for the constant $\mathcal{P}$ curves than for the constant $V$ curves when $0.5 < \beta < 1$.  This gives the shape of the contours on the left hand side of Figure \ref{fig:optmotioncontours}.  

\section{Conclusions}

   The principal results of this paper are (a) that the accuracy of the SPH algorithm for linked bodies moving in a fluid has been established, (b)  that the variation of the calculated speed and power output take  simple forms consistent with scaling relations, (c) that there is remarkable agreement between the two dimensional results and those Taylor obtained for the swimming of long narrow animals in three dimensions, and (d) the minimum power to produce a specified speed for a given gait has been calculated and forms a basis for other such calculations. 
   
    The first of these results has been obtained by resolution studies and by comparison with the results of Eldredge (obtained for massless bodies in a viscous fluid) and Kanso et al. (obtained for neutrally buoyant bodies in  inviscid fluids).  In both cases the relevant results are limits of our calculations. In the case of Eldredge we estimated his value from a series of calculations where the mass of the body was changed.  In the case of Kanso et al. the viscosity was steadily decreased so that the Reynolds number increased from 50 to 5000.  
    
    The second and third results were obtained by fixing the dimensions of the bodies and their masses but changing the frequency and amplitude of the gait.  The simulations show that the the results have a simple dependence  on frequency and amplitude which is similar to that found by Taylor (1952).  These results suggest that the drag forces on a long thin ellipse in two dimensions is similar to that on a cylinder in three dimensions.  In particular, it suggests that the drag has two additive contributions. One  varying with Reynolds number as $1/\Re^{1/2}$ and one depending on the square of the velocity relative to the fluid. We are unaware of calculations or analysis which would confirm this conjecture in detail.  It is clear however, that there will be pressure forces proportional to the square of the velocity on the bodies, and viscous forces  due to flow along and between the ellipses. 
    
     The result (d) shows that the efficiency is poor if  the linked bodies are driven with a gait amplitude which is too large or too small.  We find that the optimum performance occurs when the amplitude $\beta \sim 1.2$ or, equivalently, when the angles between the links varies between $\pm \pi/3$.
   
The formulation we have used is general and can be immediately applied to the motion of linked bodies in stratified fluids, or with a free surface, or within complex boundaries, or with more complex constraints including those where the gait depends on the positions of the bodies in the domain.  We are currently studying these problems.

\end{document}